\documentclass[twocolumn]{aastex62}
\usepackage{amsmath,amstext}
\usepackage[T1]{fontenc}
\usepackage{soul}
\usepackage[figure,figure*]{hypcap}
\shorttitle{The first identified Group III CEMP-no star in CVn I}
\shortauthors{Yoon et al.}
\begin{document}

\title{Identification of a Group III CEMP-no Star in the Dwarf Spheroidal Galaxy Canes Venatici I}

\author[0000-0002-4168-239X]{Jinmi Yoon}
\affiliation{Department of Physics and JINA Center for the Evolution of the Elements, University of Notre Dame, Notre Dame, IN 46556, USA}
\correspondingauthor{Jinmi Yoon}
\email{jinmi.yoon@nd.edu}

\author[0000-0002-9594-6143]{Devin D. Whitten}
\affiliation{Department of Physics and JINA Center for the Evolution of the Elements, University of Notre Dame, Notre Dame, IN 46556, USA}

\author[0000-0003-4573-6233]{Timothy C. Beers}
\affiliation{Department of Physics and JINA Center for the Evolution of the Elements, University of Notre Dame, Notre Dame, IN 46556, USA}

\author[0000-0001-5297-4518]{Young Sun Lee}
\affiliation{Department of Astronomy and Space Science, Chungnam National University, Daejeon 34134, Korea}

\author{Thomas Masseron}
\affiliation{
Instituto de Astrof\'{i}sica de Canarias, E-38205 La Laguna, Tenerife, Spain}
\affiliation{Departamento de Astrof\'{i}sica, Universidad de La Laguna, 38206, La Laguna, Tenerife, Spain}

\author[0000-0003-4479-1265]{Vinicius M. Placco}
\affiliation{Department of Physics and JINA Center for the Evolution of the Elements, University of Notre Dame, Notre Dame, IN 46556, USA}

\begin{abstract}
CEMP-no stars, a subclass of carbon-enhanced metal-poor (CEMP) stars, are one of the most significant stellar populations in Galactic Archaeology, because they dominate the low end of the metallicity distribution function, providing information on the early star-formation and chemical-evolution history of the Milky Way and its satellite galaxies. Here we present an analysis of low-resolution ($R \sim 1,800$) optical spectroscopy for a CEMP giant, SDSS~J132755.56+333521.7, observed with the Large Binocular Telescope (LBT), one of the brightest ($g \sim 20.5$) members of the classical dwarf spheroidal galaxy, Canes Venatici I (CVn I). Many CEMP stars discovered to date have very cool effective temperatures ($T_{\mathrm{eff}}< 4500$\,K), resulting in strong veiling by molecular carbon bands over their optical spectra at low/medium spectral resolution. We introduce a technique to mitigate the  carbon-veiling problem to obtain reliable stellar parameters, and validate this method with LBT low-resolution optical spectra of the ultra metal-poor ([Fe/H] = $-4.0$) CEMP-no dwarf, G~77-61, and seven additional very cool CEMP stars, which have published high-resolution spectroscopic parameters. We apply this technique to the LBT spectrum of SDSS~J132755.56+333521.7. We find that this star is well-described with parameters  $T_{\mathrm{eff}}=4530$\,K, log $g=$ 0.7, [Fe/H] $ = -3.38$, and absolute carbon abundance $A$(C) = 7.23, indicating that it is likely the first Group III CEMP-no star identified in CVn I.
The Group III identification of this star suggests that it is a member of the extremely metal-poor population in CVn I, which may have been accreted into its halo.
\end{abstract}


\section{Introduction}\label{introduction} 

The nature of the first generation of stars, in particular, the first-star initial mass function (FIMF) and first-star nucleosynthesis pathways, provide crucial information for understanding the first star-forming environments and early Galactic chemical evolution.
While constraining the FIMF remains a challenge, our understanding of first-star nucleosynthesis has been advanced through studies of their likely direct descendants, the so-called CEMP-no stars ([C/Fe] $\geq +0.7$ and [Ba/Fe] $\leq 0.0$\footnote{Here we adopt the relative abundance ratios as [A/B]= log $(\rm N_A/N_B)_{\star}$ - log $(\rm N_A/N_B)_{\odot}$, where $\rm N_A$ and $\rm N_B$ are the number densities of elements A and B, respectively.}), a sub-class of carbon-enhanced metal-poor stars \citep[CEMP;][]{beers2005, aoki2007, hansen2016a,placco2016b,yoon2016, aguado2018a, ezzedine2019,frebel2019, yoon2019}. 

CEMP-no stars exhibit over-abundances of carbon but sub-solar abundance ratios of neutron-capture elements. Based on a variety of studies, \citep[e.g.,][]{yong2013,aoki2018}, a substantial fraction of extremely metal-poor (EMP; [Fe/H] $<-3.0$) and more than 50\% of ultra metal-poor (UMP; [Fe/H] $<-4.0$) CEMP-no stars also exhibit over-abundances of light elements such as N, O, Na, Mg, Al, and Si, which may be a characteristic signature of first-star nucleosynthesis \citep{aoki2002, meynet2010, nomoto2013, choplin2017, aoki2018}. 

Recently, \citet{yoon2016} demonstrated that halo CEMP-no stars can be sub-divided into at least two groups, based on their morphological distinction in the Yoon-Beers $A$(C)\footnote{$A$(X)=$\log\,\epsilon$(X)=$\log\,$($N_{\rm X}/N_{\rm H}$)+12, where $N_{\rm X}$ and $N_{\rm H}$ the represent number-density fractions of the element X and hydrogen, respectively.}-[Fe/H] diagram. The Group II CEMP-no stars exhibit a strong correlation between $A$(C) and [Fe/H], indicating formation in environments where progenitor stars  simultaneously produced both carbon and iron. In contrast, the progenitors of the Group III CEMP-no stars appear to have produced carbon independently of iron.  Similar bifurcated behaviors between Group II and III CEMP-no stars are found in the $A$(Na)-$A$(C) and $A$(Mg)-$A$(C) spaces, along with the recently explored $A$(Ba)-$A$(C) space \citep{yoon2019}. These chemically distinct behaviors strongly suggest the existence of multiple nucleosynthesis pathways for the formation of CEMP-no stars. Hence, recent theoretical studies have investigated different explanations for the formations of the bifurcated CEMP-no stars, such as different cooling channels (carbon dust grains vs. silicate dust grains; \citealt{chiaki2017}), different pollution/metal-enrichment pathways  (external vs. internal enrichments; \citealt{chiaki2018, chiaki2019}), or inhomogeneous metal mixing of their birth clouds \citep{hartwig2019}.

\begin{figure*}
\centering
\includegraphics[trim = 3.0cm 5.0cm 1.50cm 2.0cm, clip, width=\textwidth]{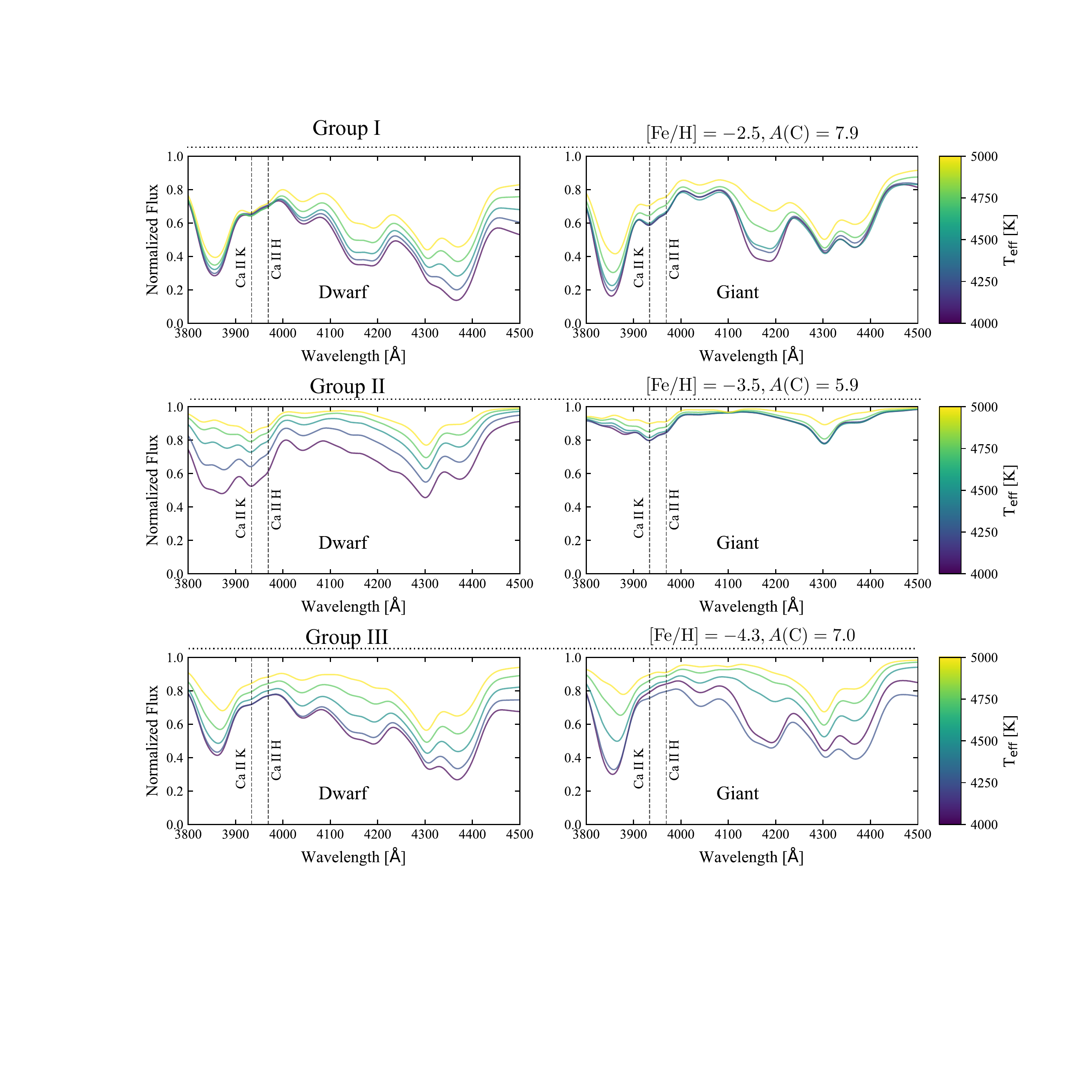}
\caption{Influence of carbon veiling in the region of the \ion{Ca}{2} H\,\&\,K lines and the CH $G$-band, for CEMP Group I (\textit{top panels}), Group II (\textit{middle panels}), and Group III (\textit{bottom panels}) archetypes, as a function of temperature, over  $4000\,{\textrm K}\leq T_{\rm eff} \leq 5000\,{\textrm K}$. Cooler atmospheres are characteristically more affected by continuum depression than those with higher effective temperatures. Surface gravities for dwarf (\textit{left panels}, $\log g=5.0$) and giant (\textit{right panels}, $\log g = [0.0, 3.5]$) classifications are also considered, for which the behavior of carbon veiling contrasts significantly. \label{fig:c_veil}}
\end{figure*}

Though some UMP CEMP-no stars are confined close to the Milky Way's plane, within 3\,kpc (\citealt{sestito2019}, Yoon et al., in prep), CEMP-no stars in the halo exhibit kinematics that suggest their dominant association with the outer-halo population of the Milky Way \citep[e.g.,][]{carollo2014,lee2017, yoon2018, lee2019}. We note that \citet{chansen2019} reported that their kinematics study of a sample of CEMP-no stars does not necessarily support this idea; their inference was based, however, on a rather small sample size ($\sim$30). The halo CEMP-no stars are likely to have been accreted from their birthplaces, dark-matter dominated low-mass mini-halos \citep[e.g.,][]{salvadori2015,amorisco2017,starkenburg2017}, that are often taken to be the site of first-galaxy formation.  The accretion origin of the halo CEMP-no stars has been supported by various studies asserting their association with satellite dwarf galaxies in the Milky Way \citep[e.g.,][]{frebel2015,starkenburg2017,spite2018}. Most recently, \cite{yoon2019} demonstrated the existence of a similar bifurcated behavior of CEMP-no Groups in the Yoon-Beers diagram among the sample of CEMP-no stars found in the ultra-faint dwarf (UFD) satellite galaxies and classical dwarf spheroidal (dSph) galaxies, providing additional strong evidence for the accretion hypothesis.

The discovery of large numbers of halo CEMP-no stars has been limited by the need for a measurement of [Ba/Fe], which requires time-consuming moderate- to high-resolution spectroscopy. \citet{yoon2016} devised an alternative approach, based on the absolute carbon abundance, $A$(C), which is readily measured at low resolution, and capable of effectively (with a success rate of $\sim$90\%) distinguishing CEMP-no stars from CEMP-$s$ stars, whose carbon and barium over-abundances are thought to originate from mass transfer from their asymptotic giant branch binary companion, which is now a faint white dwarf \citep{suda2004,herwig2005,lucatello2005,komiya2007,bisterzo2011,starkenburg2014,hansen2016b,arentsen2019, lee2019}.  In particular, in the metallicity range where these two CEMP sub-classes overlap, $-3.5 \leq$ [Fe/H] $\leq -2.0$, application of this approach opens the opportunity to identify significantly greater numbers of CEMP-no stars. 

However, there remains a challenge for the estimation of reliable stellar parameters from low- and medium-resolution spectroscopy, in particular for cooler stars ($T_{\rm eff} < 4500$\,K), due to the presence of strong molecular carbon bands throughout the optical spectral region of interest. In this work, we develop a technique to mitigate this limitation, both by assigning individual cool CEMP stars into their likely CEMP Groups in the $A$(C)-[Fe/H] diagram, and performing concurrent determinations of effective temperature, metallicity, and carbon abundance. We validate this method by deriving the stellar parameters of known very cool CEMP stars ($T_{\mathrm{eff}}< 4500$\,K) whose high-resolution spectroscopic parameters are available, including a new low-resolution spectrum of the canonical ultra metal-poor (Group III CEMP-no) dwarf carbon star, G~77-61.  

We employ this method to recently acquired low-resolution optical spectroscopic observations of a carbon giant, SDSS~J132755.56+333521.7 (hereafter, SDSS~J1327+3335), a member of the satellite dwarf galaxy Canes Venatici I (CVn I), to derive its stellar parameters. SDSS~J1327+3335 was found in close proximity to the center of CVn I, which was originally revealed as a stellar over-density in the North Galactic Cap using Sloan Digital Sky Survey Data Release 5 \citep[SDSS,][]{york2000, yanny2009}. Although the existence of this star was reported in the CVn I discovery paper \citep{zucker2006}, the poor quality of the original SDSS data, due to its faint magnitude ($g\sim20.5$), rendered it unusable for reliable stellar-parameter estimates.   

In Section~\ref{observation}, we describe the low-resolution ($R \sim$ 1,800) spectroscopic follow-up observations of G~77-61 and SDSS~J1327+3335 and additional validation stars.  We introduce our method for identifying the likely group membership and derive their stellar parameters by implementing a Bayesian Maximum Likelihood Estimation (MLE) spectral-matching procedure in Section~\ref{method}. In Section~\ref{results}, we validate this method by comparing our derived estimates for G~77-61 and other seven CEMP stars with high-resolution spectroscopic results. We then apply this method to a low-resolution spectrum of SDSS~J1327+3335. In Section~\ref{discussion}, we confirm that SDSS~J1327+3335 is a likely Group III CEMP-no star, based on its location in the $A$(C)-[Fe/H] space, and discuss its significance in terms of its host environment and accretion origin. Our conclusions are summarized in Section~\ref{conclusion}.

\section{DATA}\label{observation}
In this section, we describe the various spectroscopic data made use of in this work.

\subsection{Validation Stars}
As a validation of the classification and parameter determination methodologies, we select known cool ($T_{\rm eff} < 4500$\,K) stars, previously studied with high-resolution spectroscopy, for which the metal and carbon abundances are prime examples of their CEMP Group classifications. There is an exception\footnote{There are only two known Group III stars with 4000\,K $\leq T_{\rm eff} \leq 5000$\,K: G~77-61 ($T_{\rm eff} = 4000$\,K) and HE~1310-0536 ($T_{\rm eff} = 5000$\,K).} with a warmer Group III star with 4500\,K $<T_{\rm eff} \leq 5000$\,K. This selection includes several spectra for validation --  six stars from the Hamburg/ESO survey (HES; \citealt{Wisotzki:1996, christlieb2001}), one spectrum from the HK objective-prism survey \citep{beers1985,beers1992}, and our new low-resolution spectrum of G~77-61, as described in the next subsection. These stars include four CEMP Group I stars (HE~1305+0132, HE~2221-0453, HE~0319-0215, and HE~0017+0055),  two Group II stars (HE~1116-0634 and CS~30314-0067\footnote{\citet{yoon2016} noted that this star is located in the overlapping regions between Group II and Group III.}), and  two Group III stars (HE~1310-0536 and G~77-61). 

\subsection{Low-Resolution Spectroscopic Observations and Data Reduction}
While we used existing low-resolution spectra from both the HES survey and HK survey for validation (typical signal-to-noise ratio, SNR\footnote{Our method can be implemented for spectra of $\textrm{SNR}\sim 20 $ at 4000\AA\, and may be extendable to $\textrm{SNR} \sim 10$, and still produce acceptable results.}, estimates were $\sim 25$ at the \ion{Ca}{2} K line.), we conducted new low-resolution spectroscopic observations for G~77-61, along with our science target, SDSS~J1327+3335.
We obtained low-resolution spectroscopy of SDSS~J1327+3335 with the Multi-object Double Spectrographs \citep[MODS;][]{pogge2010} at the Large Binocular Telescope (LBT) on Mt. Graham, Arizona. For comparison, we also observed the canonical UMP dwarf carbon star G~77-61, with stellar parameters available from previous analysis of high-resolution spectroscopy \citep{plez2005}. Below we provide a description of the observations and data reduction.

The optical spectra of SDSS~J1327+3335 were obtained on May 21 and June 5 2018. We used the blue grating covering the wavelength range 3200-5800\,\AA, with a dispersion of 0.5 \AA\, pixel$^{-1}$, which provides a resolving power of $R \sim$ 1,800. The 0.6 arcsec segmented long slit was used to obtain eighteen 20-minute exposures for SDSS~J1327+3335.  The spectra were flat-fielded, bias-subtracted, and bad columns fixed using the modsCCDRed python package\footnote{modsCCDRed by R. W. Pogge, available at \url{http://www.astronomy.ohio-state.edu/MODS/Software/modsCCDRed/}}\citep{pogge2019}.  Cosmic rays were identified and removed using the L.A. Cosmic IRAF\footnote{IRAF is distributed by the National Optical Astronomy Observatory, which
is operated by the Association of Universities for Research in Astronomy
(AURA) under cooperative agreement with the National Science Foundation.} task (van Dokkum 2001).  The wavelength calibrations were carried out based on observations of Ar lamps taken during the same run with the standard LBT linelists. The sky subtraction, wavelength calibration, and one-dimensional extraction tasks were carried out using IRAF. The MODS 1 and MODS 2 spectra were co-added in the final step. Note that three of the 20-minute exposure MODS2 spectra from the June run were not co-added for the final analysis, due to a problem with their flat-field spectra. From the co-addition of fifteen spectra (total exposure
time 18,000\,s), a final SNR of $\sim$22 per resolution element at 4000\,\AA\, was achieved.



The spectra of G~77-61 were obtained on February 9, 2018 and reduced with the same procedure described above, except that we used HgAr lamp spectra for the MODS2 data for wavelength calibration. The total exposure time was 3200\,s. The resulting SNR obtained is $\sim$160 per resolution element at 4000\,\AA\footnote{Wavelength shifts of a few tenths of \AA\ were found between the MODS1 and MODS2 spectra, of an unknown origin, hence we decided not to measure radial velocity estimates. These shifts do not influence on our abundance results, because the observed spectra were corrected to the rest frame of the Balmer lines prior to all analyses conducted.}. 

\section{A Novel Method: Archetypal Classification and Spectral Matching}\label{method}

The determination of stellar parameters for cool ($T_{\mathrm{eff}} <4500$\,K), strongly carbon-enhanced stars is challenging for two primary reasons. First, large swaths of the optical spectral range can be significantly depressed by the presence of molecular carbon features and metallic lines, including regions commonly used to approximate the continua of these spectra. This issue, referred to as carbon veiling, is exhibited in Figure~\ref{fig:c_veil}, where continuum depression is seen for cooler atmospheres, causing otherwise carbon-independent features in low-resolution spectra, such as the \ion{Ca}{2} H\,\&\,K lines, to exhibit a dependence on the carbon abundance. This veiling is also sensitive to the surface gravity of the star, and behaves distinctly for each of the three CEMP Groups. Consequently, the effective temperature, metallicity, and carbon abundance of a star need to be considered simultaneously during parameter determination, and should not be regarded as independent procedures. We remind the reader that the low- and medium-resolution spectra of CEMP stars often come from large-scale surveys, which are only approximately flux-calibrated, if at all. 

The second challenge is the ambiguous nature of spectrum normalization in the presence of significant carbon veiling, especially with regards to spectral matching. It is generally not known to what extent carbon veiling is influencing the pseudo-continuum of an observed spectrum during normalization and subsequent parameter determination. Whereas the normalization of observed spectra with strongly enhanced molecular band features proceeds in a manner that might be oblivious to the degree of carbon veiling, depending on the strength of carbon bands, spectral synthesis does not; the flux is representative of the underlying stellar atmosphere. Parameter determination is therefore only possible if both the observed and synthetic spectra have addressed carbon veiling in an equivalent fashion.

We attempt to resolve these issues in the following manner. First, we employ a normalization technique designed to accommodate both the observed spectra and our library of synthetic spectra, as described below. Then, we employ a two-step procedure to derive reliable estimates of [Fe/H] and $A$(C): (1) We develop a preliminary CEMP Group classification procedure based on archetypal stellar parameters, and (2) We implement a MLE technique for spectral matching, which takes into account both metal- and carbon-abundance sensitive features during determination of effective temperature, metallicity, and carbon abundance, to derive reliable stellar parameters.

\subsection{Synthetic Spectra and Normalization}

For all synthetic spectral matching, we made use of the synthetic (1D LTE) library described in \citet{whitten2019}; a brief description follows. We implemented a grid of model atmospheres computed with the MARCS code \citep{marcs}, taking into account carbon enhancement in the atmosphere. The microturbulence velocities were assigned according to the prescription $v_{t}=-0.345 \cdot \log{g} + 2.225$, derived from a sample of high-resolution spectra \footnote{see \url{https://www.sdss.org/dr12/spectro/sspp_changes/} for details.}.
Synthetic spectra were generated from these model atmospheres using the Turbospectrum routine \citep{turbospectrum}, covering the wavelength range 3,000--5,000\,\AA. We assume the Solar abundances of \citet{asplund2009}. Updated linelists and more detailed information about the grids can be found in \citet{whitten2019}. This library was decremented to an appropriate spectral resolution ($R=2,000$), and renormalized along with the observed spectra using the Gaussian Inflection Spline Interpolation Continuum (\texttt{GISIC}) routine \footnote{\url{https://pypi.org/project/GISIC/}}, in order to enable matching with the observed spectra. This routine implements a cubic spline interpolation of continuum regions determined from inflection points in the smoothed spectrum. We note that GISIC may be a useful tool for application to other large spectral samples, in particular where automated approaches are employed.

We first interpolate within this library to produce spectra across the range $T_{\rm eff}$ = [4000, 5000]\,K, [Fe/H] = [$-4.5,-1.0$], and [C/Fe] = [$-0.5, +4.5$]. Given the effective temperature range considered in this work, we assume both dwarf and giant surface gravity classifications. For both classifications, surface gravities are assigned according to the effective temperature and metallicity, using stellar isochrones from the $Y^2$ collaboration \citep{demarque2004}.  An $\alpha$-element enhancement of [$\alpha$/Fe] = +0.4 -- consistent with the halo stars \citep{ishigaki2013} -- is assumed, with an age of 12\,Gyr. It is expected that the luminosity class is known prior to classification and parameter determination. In the case that the luminosity class is not known, the methods discussed can be used to determine the most likely classification, by comparing the optimized likelihood functions for both dwarf and giant classifications.

\begin{table}
	\caption{CEMP Group Archetype Parameters for the Halo and UFD/dSph Galaxies}             
	\label{table:archetype_parameters}      
	\centering                          
	\begin{tabular}{l c c c}        
		\hline\hline                  
		Group & [Fe/H] &[C/Fe] & $A$(C)\\  

		\hline   
        \multicolumn{4}{c}{MW Halo stars} \\
        \hline
        Group I   & $-2.5$ &  1.97& $7.9$ \\
        Group II  & $-3.5$ &  0.97 & $5.9$ \\
        Group III & $-4.3$ & 2.87 & $7.0$ \\
        \hline
        \multicolumn{4}{c}{UFDs/dSphs stars} \\
        \hline
        Group I   & $-1.5$ & 1.07 & $8.0$ \\
        Group II  & $-3.0$ & 0.87 & $6.3$ \\
        Group III & $-3.5$ & 2.37 & $7.3$ \\
        \hline        
		
		\hline                                   
	\end{tabular}
\end{table}


\subsection{Archetypal Classification}\label{section:classification}

\begin{figure}
\centering
\includegraphics[trim = 12.000cm 9.0cm 12.50cm 9.0cm, clip, width=\columnwidth, scale=0.85]{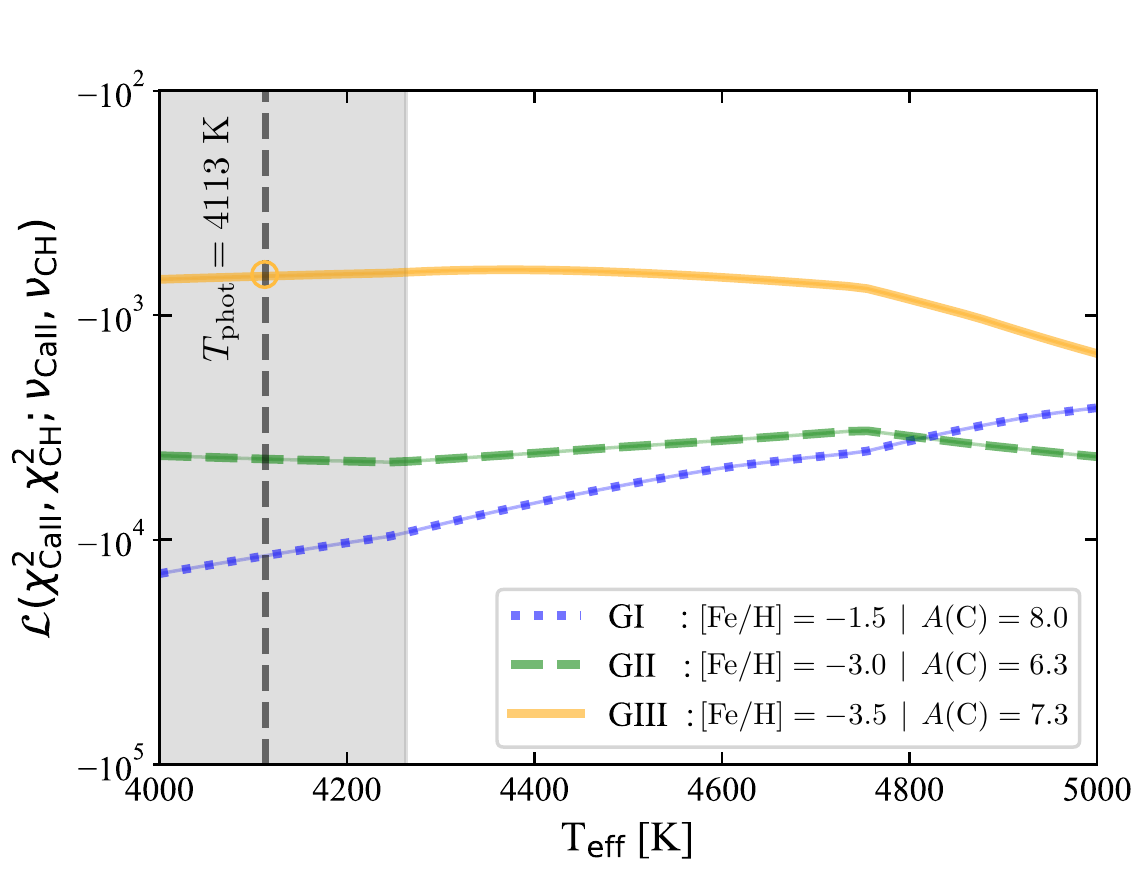}
\caption{CEMP Group archetype classification (UFD/dSph) of SDSS~J1327+3335. The log-likelihood ($\mathcal{L}$; Eq. 2) is determined across the temperature range $T_{\rm eff} =[4000, 5000]$\,K for each CEMP Group, using the [Fe/H] and $A$(C) archetype parameters, according to Table~\ref{table:archetype_parameters}. \label{fig:SDSS_class}}
\end{figure}

The distinct locations of the CEMP Groups in the $A$(C)-[Fe/H] space \citep{yoon2016} present an opportunity to determine the most likely group classification prior to parameter determination, based on comparison with archetypal parameters associated with each group, as listed in Table~\ref{table:archetype_parameters}.  The archetypal parameters were chosen based on the crude mid-points of the CEMP Group ellipses of the Yoon-Beers diagram. Thus, we referred the CEMP ellipses from Figure 1 of \citet{yoon2016} for the halo stars and Figure 2 of \citet{yoon2019} for the dwarf galaxy stars. This preliminary classification is not strictly required for the parameter determination -- described in Section~\ref{section:mcmc_method} -- as the routine is applied over the entire synthetic library range, regardless of the suggested group classification. However, the CEMP class archetype parameters are used to seed the initial values of the parameter determination, and therefore serve as a first guess of the final parameters. 

When comparing to the CEMP Group archetype parameters, it is important to consider the population to which a given star belongs. \citet{yoon2019} showed that, while the characteristic CEMP Groups are seen for stars in UFD/dSph galaxies, a slight offset exists in the [Fe/H] vs. $A$(C) behavior between the UFD/dSph and halo CEMP-no stars. This offset is likely to be associated with less dilution of the nucleosynthetic products from the progenitor stars in UFDs/dSphs, due to their lower baryonic mass reservoirs compared with the birth mini-halos of the halo stars. We therefore proceed with the UFD/dSph parameters for SDSS~J1327+3335, and utilize the halo archetypes for all validation halo stars considered in this work, according to Table~\ref{table:archetype_parameters}.

We determine the classification likelihood, on the basis of the chi-squared statistic, for both the \ion{Ca}{2} K line and CH $G$-band,

\begin{equation}\label{eq:chi}
    \chi^2(\mathbf{f}, \mathbf{E} | \xi) = \sum_i^n \frac{(f_i - E_i)^2}{(\xi E_i)^2} \hspace{0.5cm}.
\end{equation}

\noindent Here, $f_i$ and $E_i$ are the normalized fluxes of the observed and synthetic spectra, respectively, where $i$ represents each data point in the wavelength range considered, as shown Table~\ref{table:feature_sidebands}. The uncertainty used as denominator in Eq.~\ref{eq:chi} is determined as  $\xi E_i$, where $\xi$ is the inverse of the SNR, $\xi = 1/\textrm{SNR}$, for each region, corresponding to a percent uncertainty ($\xi \in (0, 1)$). The reason we introduce the inverse SNR, $\xi$, is to derive an uncertainty on the normalized flux. We discuss the formal estimation of $\xi$ in Section~\ref{section:mcmc_method}. For preliminary classification, this value is determined separately for \ion{Ca}{2} K and the CH $G$-band, using the average SNR within the sidebands listed in Table~\ref{table:feature_sidebands}.
 
\begin{table}
	\caption{Wavelength Bands used for SNR Estimation}             
	\label{table:feature_sidebands}      
	\centering                          
	\scriptsize
	\begin{tabular}{l c c c}        
		\hline\hline                  
		Feature & Blue sideband (\AA) & Line band (\AA) & Red sideband (\AA)\\  
		
		\hline   
		\hline
		\ion{Ca}{2} K  & [3884, 3923]  &  KP Eq.~\ref{eq:kp} &  [3995, 4045] \\
		CH $G$-band & [4000, 4080]  &    [4222, 4322]   &  [4440, 4500] \\
		C$_2$ (4737 \AA) & [4500, 4600] &     [4710, 4750]  &  [4760, 4820] \\
		\hline
		\hline        
		
		\hline                                   
	\end{tabular}
\end{table}

We consider the likelihood of each class as the product of the $\chi^2$-distribution probability density function for each feature: 

\begin{equation}\label{eq:likelihood}
\begin{split}
    \mathcal{L}(\chi^2_{\rm CaII}, \chi^2_{\rm CH} | \nu_{\rm CaII}, \nu_{\rm CH}) = \\  \rho(\chi^2_{\rm CaII}; \nu_{\rm CaII}) \cdot \rho(\chi^2_{\rm CH}; \nu_{\rm CH})
\end{split}
\end{equation}

\noindent Here, $\rho$ is the $\chi^2$ probability distribution, and $\nu$ is the degrees of freedom, ($n-1$), for the spectral region of interest. The product of the $\chi^2$ likelihood is used as a means of mitigating the relative difference in wavelength range between the \ion{Ca}{2} K and the CH $G$-band features.

It is important to consider the bandwidth over which the $\chi^2$ value is estimated. In general, the greater the absorption line, the larger the bandwidth should be, in order to best evaluate the characteristics of the feature. For the \ion{Ca}{2} K line, we therefore implement the band-switching scheme developed by \citet{Beers:1990a} for the KP equivalent-width estimator.

\begin{equation}\label{eq:kp}
\Delta \lambda_{\rm CaII} =
\begin{cases}
[3930.7, 3936.7]  \textrm{\AA \hspace{0.1cm} if\hspace{0.1cm} K}_{6} \le 2 \textrm{\AA},\\
[3927.7, 3939.7]    \textrm{\AA\hspace{0.1cm} if\hspace{0.1cm} K}_{6} \ge 2, \textrm{K}_{12} \le 5 \textrm{\AA}, \\
[3924.7, 3942.7] \textrm{\AA\hspace{0.1cm} if\hspace{0.1cm} K}_{12} > 5 \textrm{\AA}
\end{cases}
\end{equation}

\noindent Here, K$_{6}$ and K$_{12}$ correspond to pseudo-equivalent widths of the \ion{Ca}{2} K line,

\begin{equation}
	\textrm{K}_n = \int_{\lambda_0- n/2}^{\lambda_0 + n/2} \left(1 - f(\lambda)\right)d\lambda,
\end{equation}

\noindent 
where $\lambda_0$ is the rest-frame wavelength of the \ion{Ca}{2} K line,  3933.7\,\AA. 

We compute the likelihood, $\mathcal{L}(\chi^2_{\rm CaII}, \chi^2_{\rm CH})$, for each group archetype across the effective temperature range $T_{\rm eff}=[4000, 5000]$\,K, according to the archetype parameters listed in Table~\ref{table:archetype_parameters}. The likelihood of each classification is considered against the photometric temperature estimate, determined using implementations of the Infrared Flux Method (IRFM)\footnote{The final photometric temparature was derived by averaging temperature estimates from these three methods. We note that these methods may not be valid for some EMP/UMP CEMP stars, in particular, those with very strong carbon enhancement. However, we used this estimate only for the preliminary CEMP Group assignment, and as an input parameter for the MCMC method to derive the final value.} from \citet{Bergeat:2001}, \citet{Hernandez:2009}, and \citet{Casagrande:2010}.
As 2MASS photometry was not available for SDSS~J1327+3335, we make use of the IRFM\footnote{ \url{http://www.sdss3.org/dr10/spectro/sspp_irfm.php}} adopted for the SEGUE Stellar Parameter Pipeline \citep{lee2008a,lee2008b,allende2008}, for which we determine an effective temperature of 4113\,K. The resulting log-likelihoods ($\log{\mathcal{L}}$) of SDSS~J1327+3335 and the six validation stars are shown in Figure~\ref{fig:SDSS_class} and Appendix Figure~\ref{fig:classification}, respectively. We discuss the detailed results in Section~\ref{results}.
 
\subsection{MLE Parameter Determination}\label{section:mcmc_method}

Here we describe the method developed to produce estimates of $T_{\rm eff}$, [Fe/H], and [C/Fe]. Best-fit parameters are determined for our low-resolution spectra using maximum likelihood spectral matching, for which we explore our likelihood function by sampling over the $T_{\rm eff}$, [Fe/H], and [C/Fe] parameter space of our synthetic library using the Python module \texttt{emcee} \citep{Foreman_Mackey_2013}. This module is based on Goodman \& Weare's affine invariant Markov Chain Monte Carlo routine (MCMC; \citealt{Goodman:2010}). We emphasize that, while the preliminary CEMP Group classification from Section~\ref{section:classification} motivates the final assignment, the parameter determination is nevertheless conducted over the entire synthetic grid range: $T_{\rm eff} = [4000, 5000]$\,K, [Fe/H] = [$-4.5, -1.0$], and [C/Fe] = [$-0.5, +4.5$]. The parameter determination method is essentially independent of the Group classification, with the exception that the initial values for the MCMC are set to the archetype parameters corresponding to the stars' CEMP Group class.

We utilize three spectral features for parameter determination: the \ion{Ca}{2} K line ($\lambda_0=3933.7$ \AA), the CH $G$-band, and the C$_2$ Swan band, located at 4737\,{\AA} \citep{johnson1927, christlieb2001, cotar2019}. While the \ion{Ca}{2} K and CH $G$-band are sufficient for the preliminary classification, it was found that the CH $G$-band saturates at $A(\textrm{C}) \gtrapprox 7.5$. As this is only problematic for Group I stars, the C$_2$ Swan band\footnote{We do not expect the $G$-band and the Swan band to necessarily yield the same results, due to differences in their molecular line formation. Carbon will be used to form CO, then CH and CN, and finally C$_2$. Thus, the formation of the C$_2$ bands may differ, depending on the presence and abundance of O, N, and H \citep{ting2018, franchini2020}. 
A full analysis is beyond the scope of our present approach. In our analysis, it is apparent that the abundances from the CH and C$_2$ bands do not agree when the CH line is unsaturated (around $A$(C) $<$ 7.5). Thus, we included the Swan band as well, only when the CH band is saturated for the Group I stars.} is utilized -- in addition to the CH $G$-band -- when the pseudo-equivalent width of the CH $G$-band exceeds 40\,\AA, as determined via the line band given in Table~\ref{table:feature_sidebands}. Consideration of the relative strengths of these features -- integral to our MLE procedure -- is reminiscent of the line-depth ratio method shown to be effective for determinations of effective temperature \citep{Kovtyukh:2007}, where flux ratios of parameter-sensitive features are analogous to the difference in logarithmic likelihoods corresponding to each feature.

\subsubsection{$\chi^2$ Estimation}

We estimate the $\chi^2(\mathbf{f} | \mathbf{E}, \xi)$ of the model fit for each feature, using the line bands listed in Table~\ref{table:feature_sidebands} for the \ion{Ca}{2} K line,  CH $G$-band, and C$_2$ Swan band ($4737$\,\AA). We consider the following truncated log-$\chi^2$ probability distribution function (pdf) for each absorption feature:

\begin{equation}
    \ln{\widetilde{\rho}(\chi^2; \nu)} = (\frac{\nu}{2} -1)\ln{\chi^2} - \frac{1}{2} \chi^2.
\end{equation}

\noindent Here, $\nu$ is the degrees of freedom associated with each $\chi^2$ value. We neglect the additional terms in the $\chi^2$ pdf which depend only on $\nu$, as they are computationally cumbersome and do not influence the resulting posterior likelihood distributions. However, $\chi^2$ is a function of the assumed inverse SNR, $\xi$. We include the estimate of inverse SNR for each feature as additional parameters to be determined during optimization. For convenience, we denote the set $	\boldsymbol{\xi} = (\xi_{\rm CaII}, \xi_{\rm CH}, \xi_{\rm C2})$, 
	$\boldsymbol{\chi^2}=(\chi^2_{\rm CaII} , \chi^2_{\rm CH} , \chi^2_{\rm C2})$, and $\boldsymbol{\nu} = (\nu_{\rm CaII}, \nu_{\rm CH}, \nu_{\rm C2})$.


\begin{figure*}
\centering
\includegraphics[width=\textwidth, trim=4.25cm 0.00cm 4.25cm 0.10cm, clip]{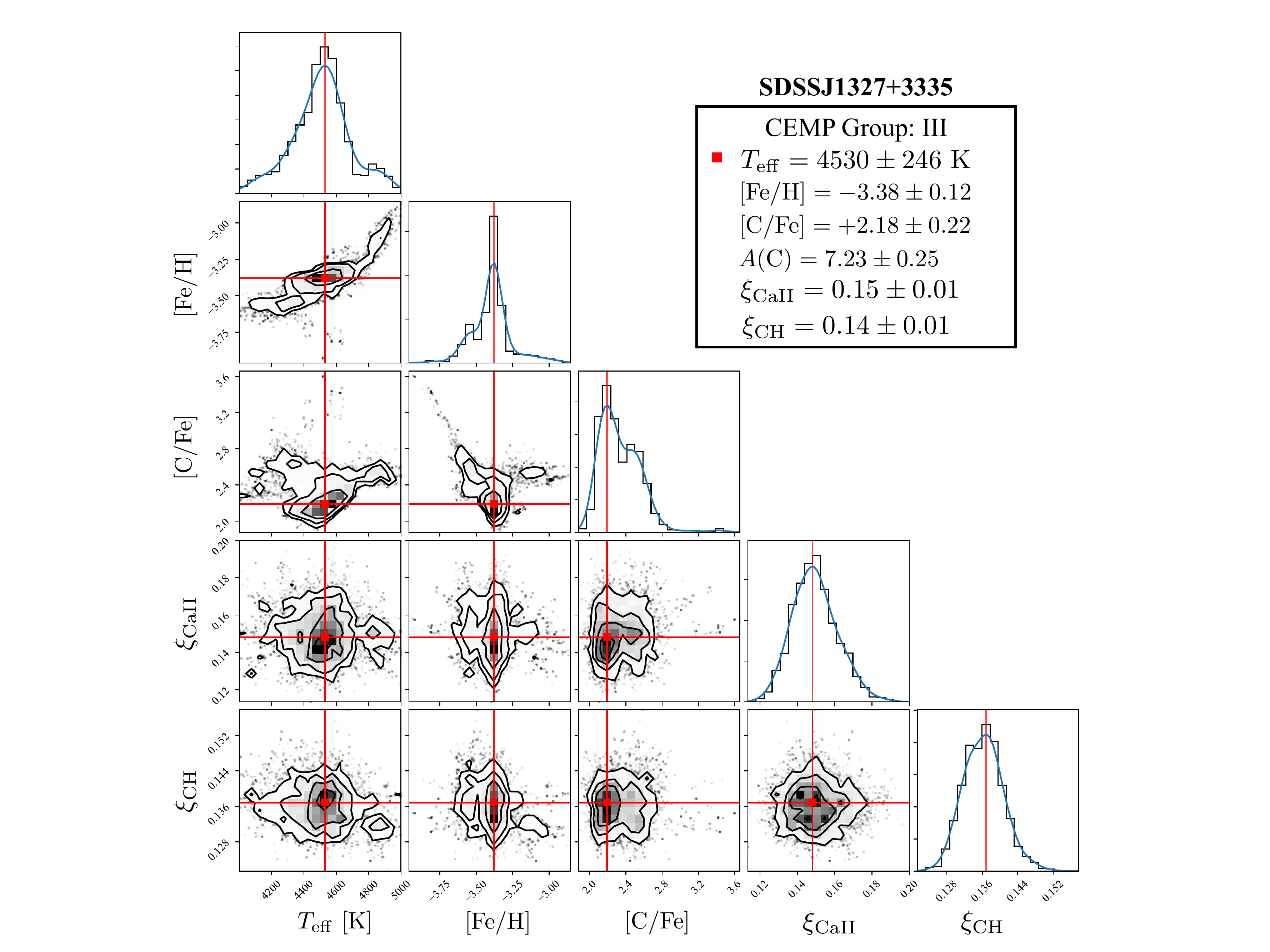}
\caption{ Posterior distributions for SDSS~J1327+3335. Best-fit parameters are determined by kernel density estimation applied to each parameter distribution. \label{fig:SDSS_corner}}
\end{figure*}


We consider the logarithm of the same likelihood used in the classification procedure, Eq.~\ref{eq:likelihood}. However, we include the C$_2$ Swan band likelihood in the event that the pseudo-equivalent width of the CH $G$-band exceeds 40\,\AA. For the set of stellar parameters, $\boldsymbol{\theta}=(T_{\rm eff}, \textrm{[Fe/H]}, \textrm{[C/Fe]}), \boldsymbol{\xi})$, the log-likelihood function is:

\begin{equation}
\begin{split}
    \ln{\mathcal{L}(\boldsymbol{\theta} | \boldsymbol{\chi^2}, \boldsymbol{\nu)}} \propto \\ 
    \ln{\widetilde{\rho}(\chi_{\rm CaII}^2; \nu_{\rm CaII})} + \ln{\widetilde{\rho}(\chi_{\rm CH}^2; \nu_{\rm CH})} + \ln{\widetilde{\rho}(\chi_{\rm C2}^2; \nu_{\rm C2})}.
\end{split}
\end{equation}

This logarithmic likelihood function is then sampled across the parameter space of the synthetic library, $T_{\rm eff} = [4000, 5000]$\,K, [Fe/H] = [$-4.5, -1.0$], and [C/Fe] = [$-0.5, +4.5$].

\begin{figure*}
	\centering
	\includegraphics[width=\textwidth, trim=3.25cm 16.50cm 3.0cm 4.0cm, clip]{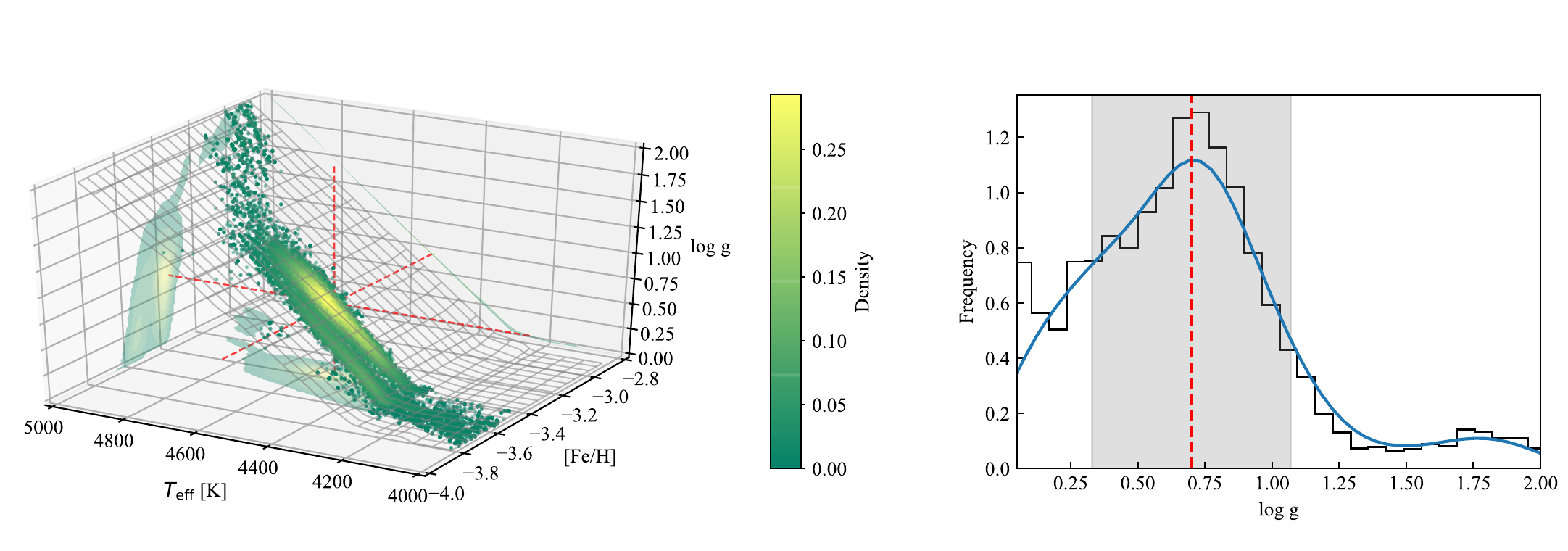}
	
	\caption{Distribution of surface gravity values for SDSS  J1327+3335. Surface gravity estimates are produced with the $Y^2$ isochrone interpolation using the posterior distributions of $T_{\rm eff}$ and [Fe/H] from the MCMC parameter determination. \textit{Left panel:} The full $T_{\rm eff}$, [Fe/H], $\log{g}$ distribution is shown, where color indicates the density of points, determined from kernel density estimation. The dashed red lines represent the maximum likelihood values. \textit{Right panel:} The distribution of $\log{g}$ values is shown, where the vertical line denotes the maximum likelihood value, $\log{g} = 0.7 \pm 0.4$. The shaded region represents the standard deviation.\label{fig:logg_posterior}}
\end{figure*}

\subsubsection{MLE Priors}

For effective temperature, we assume a Gaussian prior about the photometric temperature estimate, $T_{\rm phot}$, for which the standard deviation is set to $\sigma(T_{\rm phot}) = 250$\,K.
Priors for [Fe/H] and [C/Fe] are taken to be uniform, within the range of the synthetic library. 
The inverse SNR of each feature -- $\xi_{\rm CaII}, \xi_{\rm CH}$, and  $\xi_{\rm C2}$ -- corresponds to a percent uncertainty, $\xi \in (0,1)$. We therefore implement a beta distribution prior for each, where the mean and variance in each case is motivated by the SNR estimated from the observed spectrum, using the sidebands in Table~\ref{table:feature_sidebands} and assuming Poisson dominated noise. This is equivalent to setting the $\alpha$ and $\beta$ terms in the beta distribution. The logarithm of the prior distribution, $\rho(\boldsymbol{\theta})$, is as follows:

\begin{equation}\label{eq:main_likelihood}
\begin{split}
    \ln\rho(\boldsymbol{\theta} | \boldsymbol{\alpha}, \boldsymbol{\beta}) = \textrm{Uniform}(-4.5 \leq \textrm{[Fe/H]} \leq -1.0) \\  +\textrm{Uniform}(-0.5 \leq \textrm{[C/Fe]} \leq +4.5)  \\
    +\textrm{Uniform}(4000 \leq T_{\rm eff} \textrm{(K)} \leq 5000)  \\
     - \frac{1}{2}\ln{2 \pi \sigma(T_{\rm phot})^2} - \frac{1}{2}\left( \frac{T_{\rm eff} - T_{\rm phot}}{\sigma(T_{\rm phot})}\right)^2 \\ +
    \sum_{j=1}^{3}  \ln{\rho_{\beta}(\xi_{j} | \alpha_{j}, \beta_{j})}.
    \end{split}
\end{equation} 

\noindent
Here we use the index $j$ to denote the three spectral features of  \ion{Ca}{2} K, CH $G$-band, and C$_2$ Swan band. We remind the reader that the parameters associated with the C$_2$ Swan band, $\xi_{\rm C2}, \alpha_{\rm C2}$, and $\beta_{\rm C2}$, are only included when the pseudo-equivalent width of the CH $G$-band exceeds 40\,\AA. Otherwise, only \ion{Ca}{2} K and the CH $G$-band are considered.

We determine the parameters $T_{\rm eff}$, [Fe/H], [C/Fe], and $\boldsymbol{\xi}$, which maximize the likelihood, $\ln{\mathcal{L}}(\boldsymbol{\theta}| \boldsymbol{\chi^2}, \boldsymbol{\nu})$, by maximizing the posterior distribution of each parameter via kernel density estimation. The width of the Gaussian kernel was determined in each case by the standard deviation of the posterior distribution in question, from which the standard deviation is reported as the uncertainty estimate, $\sigma(\boldsymbol{\theta})$, where $\boldsymbol{\theta}$ represents the set of the stellar parameters ($T_{\rm eff}$, [Fe/H], [C/Fe], and $\xi$).

As seen in Table~\ref{table:validation}, the uncertainties reported for estimates of $A$(C) are always larger than the [Fe/H] estimates. This is due to the format of the spectral library being in terms of [Fe/H] and [C/Fe], which requires the $A$(C) to be determined from both. The uncertainty is then the quadrature sum of the the [Fe/H] and [C/Fe] uncertainty.

The uncertainties of our stellar abundance parameters, determined in this manner, are generally on the scale of the synthetic spectral grid resolution (0.25\,dex). In the event that they are significantly smaller, it is advised to consider the grid resolution as the primary uncertainty estimate. Uncertainties are driven largely by correlations in the stellar parameters, for instance the covariance of $T_{\rm eff}$ and [Fe/H] seen for SDSS J1327+3335 in Figure~\ref{fig:SDSS_corner}. Additionally, bi-modalities in the posterior distributions tend to increase the standard deviations reported as the uncertainties. In such cases, care should be given to the selection of the appropriate mode in the posterior distribution, which maximizes the likelihood. Bi-modality was seen to be common in the instances where the inverse signal-to-noise, $\xi$, was incorrectly determined for one or more of the spectral features. This can lead to mismanagement and over-prioritization of the features with the smallest $\xi$. For this reason, values of $\xi$ that are motivated by preliminary knowledge of the spectral signal-to-noise are typically preferred over an automated determination of the optimal $\xi$ values which maximize the likelihood function.

Surface gravity estimates are assigned via the $Y^2$ isochrone interpolation, using the values of $T_{\rm eff}$ and [Fe/H], determined from the maximum likelihood spectral matching, along with the luminosity class. To estimate the uncertainty in the surface gravity, we iteratively sample the posterior distributions of $T_{\rm eff}$ and [Fe/H] to build a distribution of gravity estimates.

\section{Results}\label{results}
\subsection{Validation of the Methodology}
We remind the reader of the two-step methodology -- the assignment of likely group classification using the archetype parameters and the parameter determination using the MCMC technique. 
The results of the CEMP Group likelihoods using the archetype parameters for the validation stars 
are shown in Appendix Figure~\ref{fig:classification}. 
These likelihood figures include the corresponding photometric temperature estimate (black-dashed line) and its uncertainty of 
$\pm 150$\,K (shaded region). Our classification technique indeed well-predicts their original group classification assigned in \citet{yoon2016} for all of the validation stars. We note that the Group II likelihood was exceedingly low for all Group I validation stars, and thus we excluded them from the figure, in order to prioritize discrimination between Group I and III classification.  

\begin{table*} 
\caption{ CEMP Group Validation Stars}             
	\label{table:validation}      
	\centering                          
	\tiny
	\begin{tabular}{l | c c c c | C C C C l}        
		\hline\hline                  
		Identifier & $T_{\rm eff}$(K) & $\log{g}$ & [Fe/H] & $A$(C)\tablenotemark{a}& $T_{\textrm{eff}}$ \textrm{(K)} & $\log{g}$ & \textrm{[Fe/H]} & $A$\textrm{(C)}\tablenotemark{a}  & Reference \\  
        \hline      
        & \multicolumn{4}{c}{This work} & \multicolumn{5}{c}{Literature values from high-resolution spectroscopy}\\
        \hline
        \multicolumn{10}{c}{Group I} \\
		\hline   
        HE~1305+0132 & $4496 \pm 130$ & $0.66 \pm 0.38$ & $-2.90 \pm 0.11$ & $ 8.28 \pm 0.19$ & 4462\pm 100 & 0.80 \pm 0.30 & -2.55\pm 0.50 & 8.57\pm 0.11 & \citet{schuler2007}\\
        
        HE~2221-0453 & $4514 \pm 170$ & $0.91 \pm 0.42$ &  $-2.48\pm 0.11$ & $8.34 \pm 0.12$ & 4400\tablenotemark{b} & 0.40\tablenotemark{b} & -2.27\pm 0.31 & 8.00\pm 0.31 & \citet{aoki2007} \\
        
        HE~0319-0215 & $4439 \pm 299$ & $0.51 \pm 0.60$ & $-2.89\pm 0.25$ & $8.15 \pm 0.30$ & 4448\tablenotemark{b} & 0.62\tablenotemark{b} & -2.30\tablenotemark{b} & 8.13\tablenotemark{b} & \citet{hansen2016b}\tablenotemark{c} \\
        
        HE~0017+0055 & $4370 \pm 48$ & $0.25 \pm 0.13$ & $-3.36 \pm 0.17$ & $7.45 \pm 0.28$ & 4146\tablenotemark{b} & 0.41\tablenotemark{b} & -2.80\tablenotemark{b} & 7.62\tablenotemark{b} & \citet{hansen2016b}\tablenotemark{c} \\
         \hline      
        \multicolumn{10}{c}{Group II} \\
        \hline
        HE~1116-0634 & $4722 \pm 87$ & $1.24 \pm 0.26$ &$-3.32 \pm 0.06$ & $5.16 \pm 0.09$ & 4400\tablenotemark{b,d} & 0.1\tablenotemark{b} & $-3.73$ & 4.78\pm 0.20 & \citet{hollek2011} \\
        CS~30314-0067 & $4141 \pm 287$ &  $-0.20 \pm0.51$\tablenotemark{g}&$-2.71 \pm 0.04$ & $6.69 \pm 0.09$ &4400\pm 100& 0.7\pm 0.3 & -2.85 \pm0.18& 6.20\pm0.18& \citet{aoki2002b} \\
        CS~30314-0067 &  &  & & &  4320\pm 12\tablenotemark{d}& 0.50\pm 0.10 & -3.01 \pm0.06\tablenotemark{e} & 6.80\tablenotemark{f}& \citet{roederer2014b} \\
        \hline
        \multicolumn{10}{c}{Group III} \\
        \hline
        HE~1310-0536 & $4904 \pm 152$ &  $1.66 \pm 0.36$&$-4.20 \pm 0.22$ & $6.69  \pm 0.32$ & 5000\pm100 & 1.9\pm0.30 & -4.15\pm0.30 & 6.64\pm0.23 & \citet{hansen2015a} \\
        G~77-61 & $4174 \pm 120$ & $5.07 \pm 0.15$\tablenotemark{g}& $-4.36 \pm 0.28$ & $7.17 \pm 0.35$ & $4000\pm200$ & 5.05\tablenotemark{b} & $ -4.00 \pm 0.15 $ & $7.0 \pm0.1$ & \citet{plez2005} \\
		
		\hline                                   
	\end{tabular}
\tablenotetext{a}{Reported value from the reference, not evolution-corrected.}
\tablenotetext{b}{Uncertainty was not reported.}
\tablenotetext{c}{\citet{hansen2016b} observed the star with the FIES spectrograph at the 2.5\,m Nordic Optical Telescope. The resolving power of each spectrum is $R \sim$ 46,000, and the average SNR is $\sim$10, thus they co-added multiple spectra of a given star to improve the SNR. }
\tablenotetext{d}{Originally reported spectroscopic temperature estimate.}
\tablenotetext{e}{[M/H], the originally reported metallicity, was derived by using only eight \ion{Fe}{2} lines. We note that other typical metal-poor studies adopt metallicity based on \ion{Fe}{1} lines. When considering 91 \ion{Fe}{1} lines of this star, the metallicity, [Fe/H], is $-3.31 \pm 0.06$.
\tablenotetext{f}{Value derived using the 2017 version of MOOG (I. Roederer 2019, private communication).}
\tablenotetext{g}{Represents values determined from $Y^2$ isochrone interpolation, fits were performed within $\log{g} = [0.0, 5.0]$, where values above or below this range were assigned $\log{g}=5.0$ or $\log{g}=0.0$, respectively.}
}

\end{table*}

\begin{figure*}
\centering
\includegraphics[width=\textwidth, trim=6.00cm 10.00cm 6.0cm 6.0cm, clip]{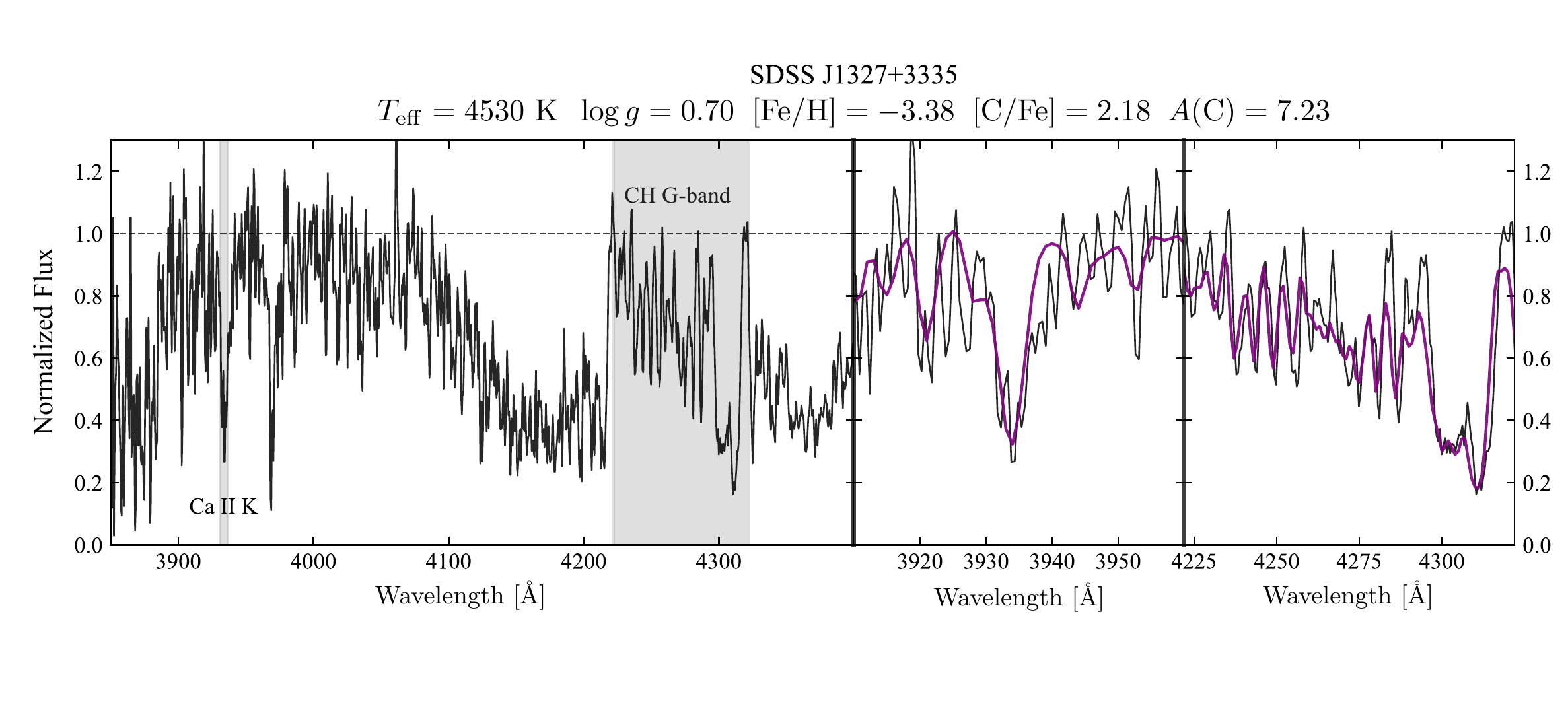}

\caption{Normalized spectra of SDSS~J1327+3335. The left-side panels represent the observed spectra (black lines) carried out with 
LBT/MODS. The gray shaded areas represent the wavelength regions of the \ion{Ca}{2} K line and the CH $G$-band, considered in the maximum likelihood estimation. The right panels shown are close-ups of the shaded regions in the left panels, with the best-fit synthetic spectra superposed (purple lines). 
\label{fig:SDSS_spectra}}
\end{figure*}

Stellar parameters $T_{\rm eff}$, [Fe/H], and [C/Fe] were determined along with the corresponding $A$(C) value for each of the validation stars using the maximum likelihood method spectral-matching procedure, as outlined in Section~\ref{section:mcmc_method}. The resulting parameters, including surface gravity, are listed in Table~\ref{table:validation}, along with those determined from previous high-resolution spectroscopic studies for comparison. The corresponding model fits are shown in Appendix Figures~\ref{fig:GI_fits} and \ref{fig:GII_GIII_fits}. 

In general, our estimates from this technique are reasonably consistent with the high-resolution spectroscopic values for the validation stars. A few exceptions are explained in more detail below.

Determinations of effective temperature agree with previous estimates within $\pm 200$\,K, which is a typical observational error, with the exclusion of HE~1116-0634, which differs from the \citet{hollek2011} estimate by $+322$\,K. However, \citet{hollek2011} adopted a "pseudo-spectroscopic" temperature for HE~1116-0634, which was obtained by applying their mean systematic offset between their spectroscopic and photometric estimates, $-225$\,K, from its photometric estimate. Their photometric temperature, 4625\,K, is consistent with our estimate (4722\,K) within the observational error. Estimates of metallicity and carbon abundance ratios for most of the validation stars generally agree within $\pm 0.4$\,dex. However, two of the Group I validation stars, HE~0319-0215 and HE~0017+0015, studied by \citet{hansen2016b}, appear to have metallicity over-estimated by $\sim$ $+0.6$\,dex, while the $A$(C) of these two stars are consistent, which is reasonable, considering the similar temperature estimates between our results and theirs. We note that these two stars were observed for radial-velocity monitoring, thus the average SNR is about 10, resulting in co-addition of the multiple spectra to improve their SNR. Thus, the \citet{hansen2016b} values are not necessarily better estimates than our results.  For CS~30314-0067, there are two high-resolution spectroscopic studies by \citet{aoki2002b} and \citet{roederer2014b}. While both of the results are reasonably consistent with ours, the $A$(C) result of \citet{aoki2002b} appears to be under-estimated by $\sim$ $-0.5$\,dex compared to our result. Since the uncertainty estimate in the surface gravity is driven entirely by the uncertainty in $T_{\rm eff}$ and [Fe/H], large uncertainties in $T_{\rm eff}$ can result in correspondingly large uncertainties in $\log{g}$. This is seen particularly for HE~0319-0215 and CS~30314-0067 in Table~\ref{table:validation}, both of which have $T_{\rm eff}$ uncertainties in excess of 250\,K and consequently $\log{g}$ uncertainties over 0.5\,dex. In the event that the photometric temperature is better constrained, either by including additional temperature calibrations or superior photometric estimates, the effective temperature determined during spectral matching can be better constrained, and thus the surface gravity uncertainty is reduced by extension.



\subsection{Application to SDSS~J1327+3335}
We apply the same methodology to our science target, SDSS~J1327+3335, taking UFDs/dSphs archetype parameters for initial Group classification. 
There is a clear preference for CEMP Group III classification across the entire effective temperature range $T_{\rm eff} = [4000, 5000]$\,K as seen in Figure~\ref{fig:SDSS_class}. The photometric temperature estimate of 4113\,K for SDSS~J1327+3335 confirms that this star is a Group III star.

The posterior distributions for $T_{\rm eff}$, [Fe/H], and [C/Fe], determined from MCMC maximum likelihood spectral matching, are shown in Figure~\ref{fig:SDSS_corner}. We determine the optimal parameters to be: $T_{\rm eff} = 4530 \pm 145$\,K, [Fe/H] $ =-3.38 \pm 0.07$, [C/Fe] $ = +2.18 \pm 0.22$, corresponding to $A$(C)$ = 7.23 \pm 0.23$. We note that this temperature is higher than the photometric estimate obtained from $g-i$ photometry, 4113\,K. However, photometric temperatures for carbon-enhanced stars -- particularly those produced from bluer filters -- are quite often underestimated, due to the strong carbon bands across the optical spectrum.
An effective temperature and metallicity of $T_{\rm eff} = 4530$\,K and [Fe/H] = $-3.38$ corresponds to a surface gravity of $\log{g}=0.7$, determined by the $Y^2$ isochrone interpolation. We estimate the uncertainty in the surface gravity determination by sampling the posterior distributions of $T_{\rm eff}$ and [Fe/H], the result of which is shown in Figure~\ref{fig:logg_posterior}. The scatter in surface gravity is largely driven by the scatter in the effective temperature, resulting in an estimate of $\log{g} = 0.7 \pm 0.4$. 
The best-fit normalized synthetic spectrum of SDSS~J1327+3335 is shown in Figure~\ref{fig:SDSS_spectra}.

\begin{figure*}
\centering
\includegraphics[trim={2.0cm 0.0cm 2.2cm 0.0cm}, clip,width=\textwidth]{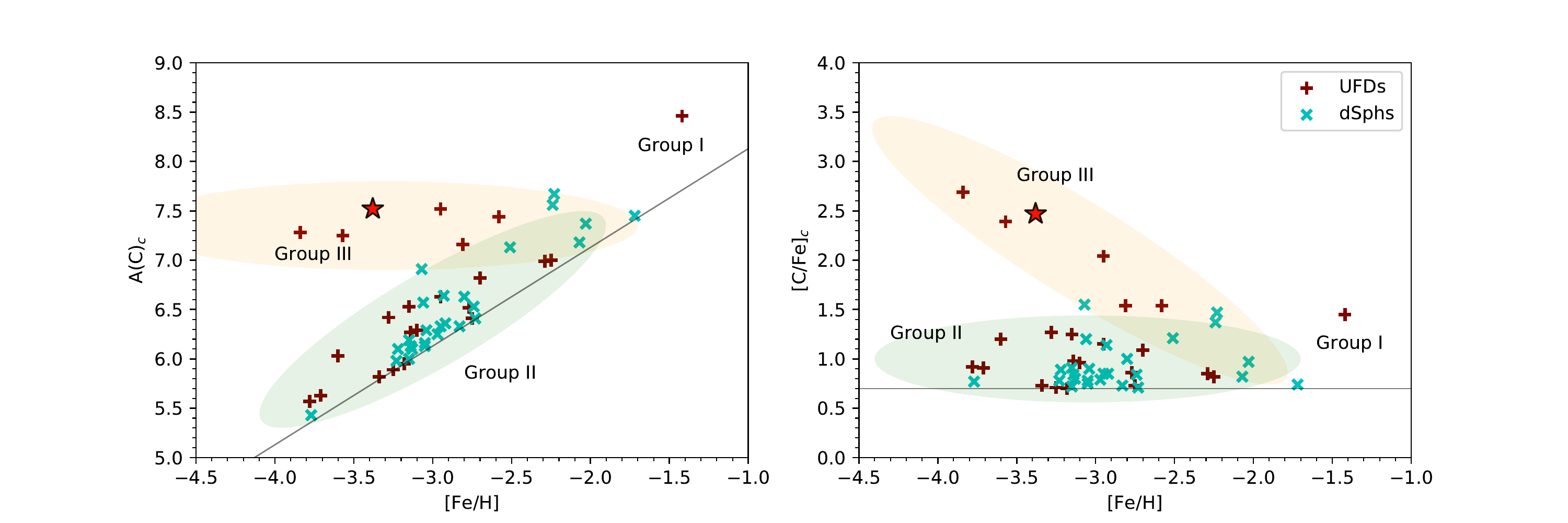}
\caption{The CEMP Group morphology for dwarf galaxies (adapted from \citealt{yoon2019}). The red star represents the location of SDSS~J1327+3335. The cyan 'x's and maroon '+'s represent CEMP stars from the dSphs and the UFDs, respectively. The orange and green ellipses represent the suggested CEMP Groups III and II, respectively. The gray lines represent [C/Fe]$_c$ = +0.7. We note that carbon abundances in these figures are evolution-corrected.
\label{ybd}}
\end{figure*}

We can confirm the luminosity class of SDSS~J1327+3335 as a giant based on its null parallax ($\pi = -0.3605 \pm	0.4190$ mas) and small proper motion ($\mu_{\rm ra} = -0.567 \pm 0.568$ mas yr$^{-1}$, $\mu_{\rm dec} = 0.471 \pm 0.300$ mas yr$^{-1}$) from {\it Gaia} Data Release 2 \citep[][]{gaiadr12016,gaiadr22018,arenou2018}. We note that this star also possesses a radial velocity of 36 $\pm$ 20\,km~s$^{-1}$, commensurate with that of other members of the CVn I dwarf satellite \citep{zucker2006}.  

Since SDSS~J1327+3335 is a late-type giant, it is clear that this star has gone through at least one dredge-up episode during its evolution. We attempt to recover the original carbon abundance from application of the carbon evolutionary correction calculator\footnote{\url{https://vplacco.pythonanywhere.com/}} of \citet{placco2014c}, obtaining a change in the carbon of $\Delta\textrm{[C/Fe]} = +0.28$\,dex. The final corrected carbon abundance is $A$(C)$_c$ = 7.51 ([C/Fe]$_c= +2.46$). The rest of our discussion below refers to these corrected abundances.

\section{Discussion}\label{discussion}
CVn I is a distant, faint dwarf galaxy, with a heliocentric distance of $\sim$220 kpc \citep{zucker2006}. Its absolute magnitude, M$_{\rm V} \sim -7.9$, makes it a galaxy similar to the Draco and Ursa Minor dwarf spheroidals, but its half-light radius, $\sim$ 550 pc, is larger than the others. Therefore, various studies have noted that CVn I is likely a dSph, rather than a UFD galaxy \citep[e.g.,][]{simon2019}. The likely CVn I membership of SDSS~J1327+3335 is not only based on its spatial location, but also its heliocentric radial velocity (consistent with the galaxy), in addition to the astrometric results noted above. However, since this star is very faint ($g \sim$ 20.5) and strongly carbon enhanced, a metallicity  determination was not available ($\textrm{SNR} < 2$ at 4000\,{\AA} and median SNR $ \sim$ 7 over the entire wavelength region of its SDSS spectrum). Even though there have been spectroscopic follow-up observations \citep{ibata2006,martin2007,simon2007,kirby2010,francois2016} from which metallicity and [$\alpha$/Fe] abundances for many stars in CVn I (mean metallicity is $-1.98 \pm0.01$ from \citealt{kirby2011a}) have been obtained, carbon-abundance estimates of the stars in this galaxy, including SDSS~J1327+3335, have not been previously reported.

Figure~\ref{ybd} shows the Yoon-Beers diagram for CEMP stars found in dwarf satellite galaxies. We used the same sample of stars as \citet{yoon2019}, which includes both dSphs (Draco, Sextans, Sculptor, and Ursa Minor) and several UFDs (Bootes I, Reticulum II, Segue I, Tucana II, Pisces II, and Ursa Major II). Note that we did not draw an ellipse for Group I, because there is only one Group I star (from Segue I). 

The location of SDSS~J1327+3335 in the CEMP Group morphology shown in Figure~\ref{ybd} indicates that it is a CEMP Group III star. Confirmation of its CEMP-no status with a Ba abundance measurement, e.g., from higher-resolution data, is not yet available\footnote{The carbon-veiling problem is prevalent over the entire spectrum of SDSS~J1327+3335, which is not flux-calibrated, resulting in difficulties with the identification of the proper continuum. This prevents reliable spectral synthesis analysis of other important elemental abundances, such as Ba, for confirming this star's nucleosynthetic origin based on the data in hand. In particular, due to quite strong N enhancement in the CN bands, even the strongest \ion{Ba}{2} line at 4554\AA\, suffers from severe blending with the CN lines. We note that \citet{norfolk2019} were able to carry out such an analysis for Ba from their approximately flux-calibrated, low-resolution LAMOST data by assuming that deviations in flux at 4554\AA\, is solely from the Ba enhancement. Since they did not report carbon measurements, it is not clear how to evaluate their synthesis of Ba for very cool stars with strong carbon enhancement such as our stars, which appears to be a minority of their sample.}, but our clasification appears quite likely, based on its $A$(C) and metallicity. 

The Yoon-Beers diagram of the halo CEMP stars reveals a strong correlation between the Group III CEMP stars and CEMP-no stars, and the association of the Group I stars with the CEMP-$s$ stars, with the exception of ``anomalous'' CEMP-no stars in the Group I region \citep{yoon2016}. This distinction is also  manifested in Figure 3 (the $A$(Ba)-$A$(C) diagram) of \citet{yoon2019}, which clearly shows that Group III CEMP-no stars are well-separated ( $\gtrsim+3.0$~dex from the mean $A$(Ba)) from the CEMP-$s$ stars. Hence, the characteristic CEMP Group morphology in the $A$(C)-[Fe/H] space can be used for a ``first-approximation'' diagnostic for identifying the likely nucleosynthetic origin of CEMP stars. There are several known EMP and many very metal-poor stars in the CVn I galaxy \citep{kirby2010}, thus we expect to find more CEMP-no stars in this system once measurements of carbon abundance are completed (E. Kirby, priv. comm.) 

According to \citet{yoon2019}, the classical dSph galaxies appear to possess only Group II CEMP-no stars, which may result from several causes:  1) The characteristically higher $A$(C) associated with Group III CEMP-no stars may have been diluted by the larger baryonic masses associated with dSph galaxies compared to UFD galaxies. 2) The additional production of iron associated with prolonged star-formation histories in dSphs compared to UFDs \citep[e.g.,][]{tolstoy2009,salvadori2015,debennassuti2017} may even reach levels where individual stars are not recognized as CEMP stars ([C/Fe] $< +0.7$). 3) A different class of nucleosynthetic origins such as spinstars or faint supernovae \citep{umeda2003,umeda2005, meynet2006,meynet2010,nomoto2013,chiappini2013,tominaga2014,maeder2015, frischknecht2012,frischknecht2016,yoon2016, choplin2017}. 4) Differences in the original pollution pathways (internal vs. external pollution, e.g., \citealt{smith2015,chiaki2018,chiaki2019}). 5) Differences in the available cooling agents (Group II: silicate-grain cooling vs. Group III: carbon-grain cooling; see \citealt{chiaki2017}) of the natal clouds could result in predominance of the Group II CEMP-no stars in the dSphs.

Thus, the Group III CEMP-no status of SDSS~J1327+3335 is intriguing, and may indicate that CVn I might have had an unusual star-formation history compared to other dSphs. Indeed, based on both Keck/DEIMOS spectroscopic and deep LBT photometric observations \citep{ibata2006, martin2007,kuehn2008, martin2008}, CVn I has been claimed to host two distinct stellar populations -- an extended metal-poor population ($-2.5 <$ [Fe/H] $< -2.0$) with hot kinematics
and a more metal-rich population ($-2.0 <$ [Fe/H] $<-1.5$) with a near-zero velocity dispersion, concentrated on its center, although this dichotomy has been challenged by a kinematic study using more than 200 stars \citep{simon2007}. Perhaps SDSS~J1327+3335 is a member of the extremely metal-poor population in CVn I, which might had been accreted into its halo.  Identification of additional Group II and III CEMP-no stars in this system should enable better understanding of the chemical evolution and accretion history of the CVn I galaxy. 

\section{Conclusions}\label{conclusion}
We have presented an analysis of low-resolution optical spectroscopy of SDSS~J1327+3335 and G~77-61, taken with the LBT MODS spectrographs, and developed a novel methodology to analyze such challenging cool (T$_{\rm eff}< 4500$\,K) CEMP stars. We identified the star SDSS~J1327+3335 as the first likely Group III CEMP-no ([Fe/H] $ = -3.38$, [C/Fe]$_c = +2.46$, and $A\mathrm{(C)}_c=7.51$) star in the CVn I dwarf satellite galaxy, using our archetypal classification - parameter determination methodology based on maximum likelihood spectral matching. This procedure was validated for each CEMP Group using spectra from the Hamburg/ESO survey, in addition to CS 30314-0067 and G~77-61, a well-known dwarf carbon star, all of which have published high-resolution analyses. The Group III CEMP-no classification for SDSS~J1327+3335 appears to be unusual among CEMP-no stars from the dSphs, which are predominantly associated with Group II stars \citep{yoon2019}. The apparently complex star-formation history of this galaxy may be responsible. The association of CEMP-no groups with a particular nucleosynthetic origin (and/or accretion origin) will provide information on both the chemodynamical assembly histories of individual dwarf galaxies and the halo system of the Milky Way.

We plan to apply our methodology to mitigate the effects of strong molecular carbon veiling, which complicates identification of the continuum around the region of the Ca H\,\&\,K lines and the CH $G$-band for other cool CEMP EMP/UMP candidates from: 1) our ongoing "Best and Farthest" survey \citep{yoon2018b}, observing with LBT/MODS and Gemini/GMOS, 2) numerous other cool CEMP stars with strong carbon veiling observed during the course of follow-up spectroscopy over the past few decades of metal-poor candidates from the HK survey (e.g., \citealt{beers1992}), 3) the list of CEMP candidates provided by \citet{christlieb2008}, and 4) very cool CEMP stars from the low-resolution spectroscopic surveys such as the SDSS, the AAOmega Evolution of Galactic Structure (AEGIS) survey \citep{yoon2018}, and the Large
Sky Area Multi-Object Fiber Spectroscopic Telescope survey \citep[LAMOST; ][]{cui2012}. These projects will also provide more validation stars and opportunities to improve the accuracy of our approach. This methodology can be widely applicable to numerous data from the future large moderate-resolution spectroscopic surveys such as the Dark Energy Spectroscopic Instrument(DESI; \citealt{desi2019}) survey,  the William Herschel Telescope Enhanced Area Velocity
Explorer (WEAVE; \citealt{dalton2018}) and 4MOST \citep{dejong2019}.  These efforts will allow us not only to expedite the discovery process of the most metal-poor stars, but also to calculate frequencies of CEMP Groups separately, and, in turn, provide insights regarding the shape of the FIMF.

We are currently preparing an open source Python package of our new methodology for public use. In the near future, we also plan to extend our synthetic spectral grid to include CEMP stars with even lower effective temperatures (to $T_{\rm eff}$ = 3500\,K), in order to better address cooler stars than included in our present grid. Once these grids are available, the versatility of our methodology will extend to probe/constrain the low end of IMF of Population II stars through application to the numerous cool dwarf and giant carbon stars known, and allow us to understand the transition from the FIMF to the current-day IMF.

\acknowledgments
We thank the anonymous referee for useful insights and a constructive report, which led to significant improvement
of this work. We would also like to thank I. Roederer for his feedback on our manuscript and recalculation of $A$(C) using the 2017 version of MOOG for CS~30314-0067. J.Y., D.D.W., T.C.B, and V.M.P. acknowledge partial support
from grant PHY 14-30152; Physics Frontier Center/JINA Center for the Evolution of the Elements (JINA-CEE), awarded by the US National Science Foundation. Y.S.L. acknowledges support from the National Research Foundation (NRF) of Korea grant funded by the Ministry of Science and ICT (No.2017R1A5A1070354 and NRF-2018R1A2B6003961). This paper used data obtained with the LBT MODS spectrographs built with funding from NSF grant AST-9987045 and the NSF Telescope System Instrumentation Program (TSIP), with additional funds from the Ohio Board of Regents and the Ohio State University Office of Research. The LBT is an international collaboration among institutions in the United States, Italy and Germany. LBT Corporation partners are: The University of Arizona on behalf of the Arizona university system; Istituto Nazionale di Astrofisica, Italy; LBT Beteiligungsgesellschaft, Germany, representing the Max-Planck Society, the Astrophysical Institute Potsdam, and Heidelberg University; The Ohio State University, and The Research Corporation, on behalf of The University of Notre Dame, University of Minnesota and University of Virginia. 
This research also made use of NASA's Astrophysics Data System, the SIMBAD astronomical database, operated at CDS, Strasbourg, France, and the SAGA database \citep[\url{http://sagadatabase.jp},][]{suda2008,yamada2013}. This work has made use of data from the European Space Agency (ESA) mission {\it Gaia} (\url{https://www.cosmos.esa.int/gaia}), processed by
the {\it Gaia} Data Processing and Analysis Consortium (DPAC, \url{https://www.cosmos.esa.int/web/gaia/dpac/consortium}). Funding for the DPAC has been provided by national institutions, in particular the institutions participating in the {\it Gaia} Multilateral Agreement.

\software{astropy \citep{astropy2013}, carbon evolutionary correction  calculator (\citealt{placco2014c}), emcee \citep{Foreman_Mackey_2013}, \texttt{GISIC} (\url{https://pypi.org/project/GISIC/}), modsCCDRed \citep{pogge2019} mplotlib \citep{hunter2007}, numpy \citep{numpy}    }

\bibliography{bibliography}
\appendix
\section{Fitting Results for the Validation Stars}
\begin{figure*}
	\centering
	\includegraphics[width=15cm,height=19.5cm, trim=2.0cm 8.00cm 2.0cm 0.10cm, clip]{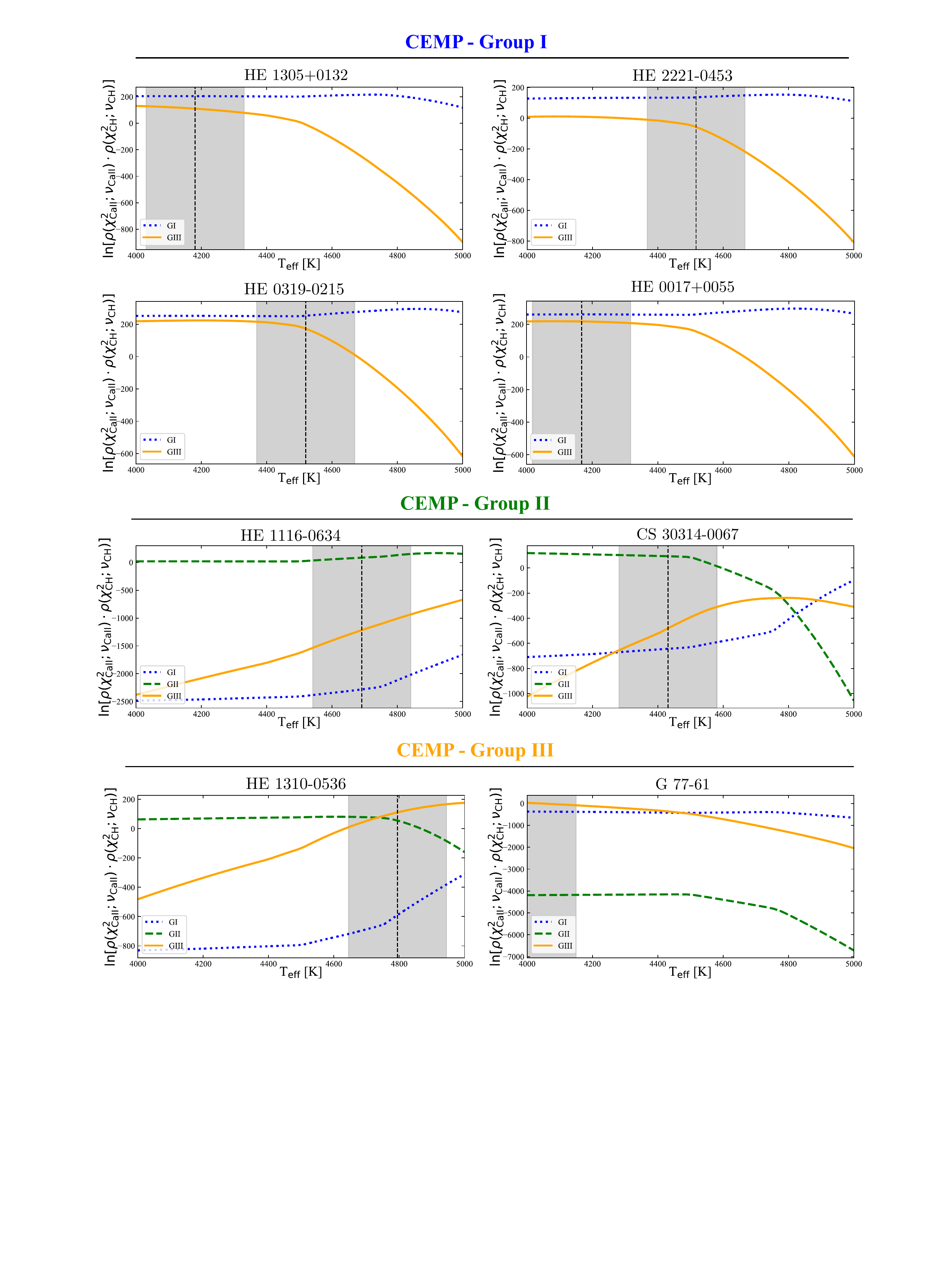}
	\caption{CEMP Group likelihoods ($\ln{\mathcal{L}}$, Eq.~\ref{eq:likelihood}) for the validation stars. Likelihoods are based on the CEMP Group archetype parameters in Table~\ref{table:archetype_parameters}. The vertical line represents the photometric temperature estimate, with $\pm 150$\,K shading.
		\label{fig:classification}}
\end{figure*}

\begin{figure*}
\centering
\includegraphics[width=\textwidth, trim=7.0cm 0.50cm 7.0cm 0.0cm, clip]{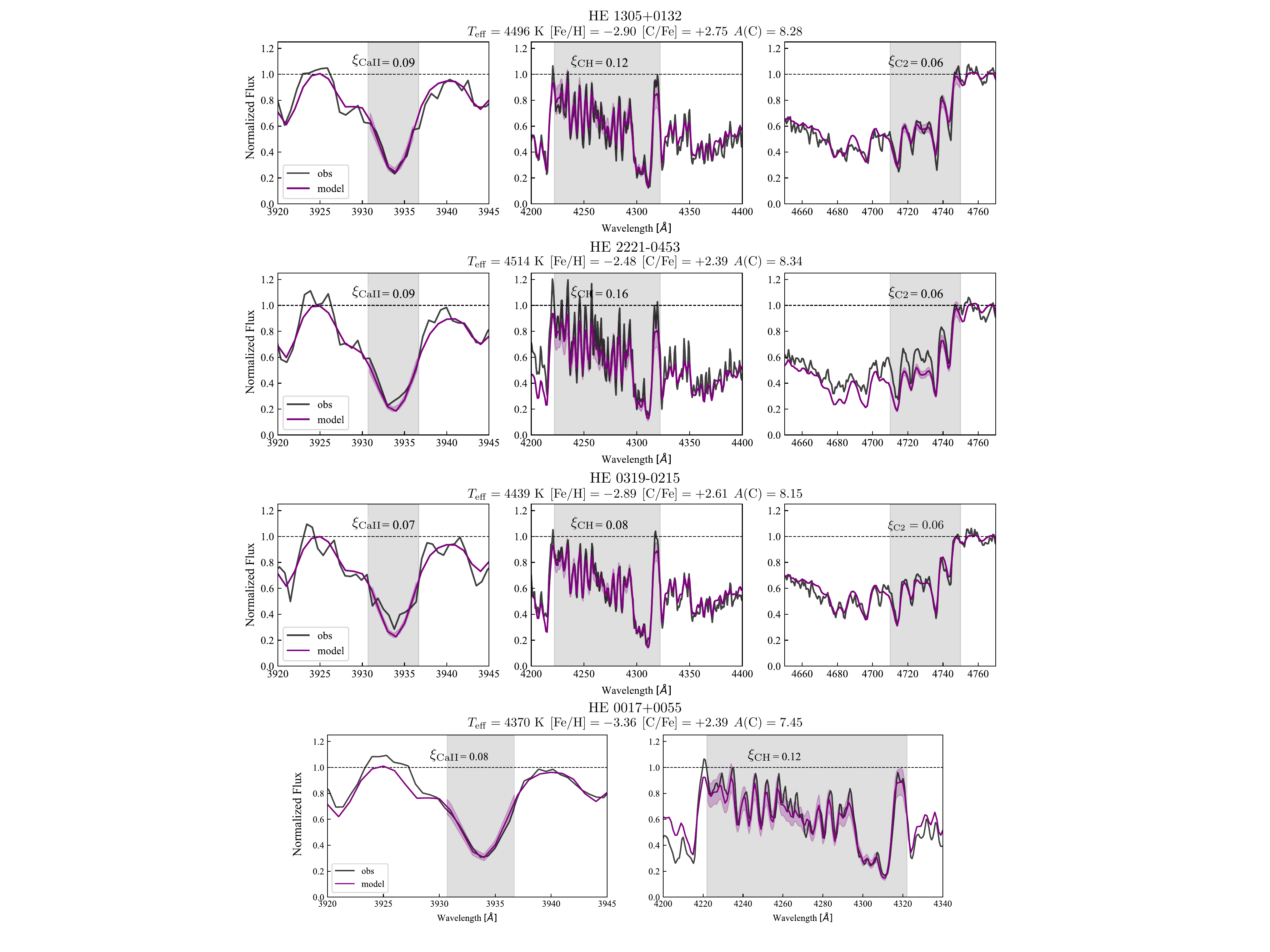}
\caption{MCMC MLE Model fits (purple) of the \ion{Ca}{2} K \textit{(left column)}, CH $G$-band \textit{(center column)}, and C$_2$ Swan band \textit{(right column)}, for the Group I validation stars: HE~1305+0132, HE~2221-0453, HE~0319-0215, and HE~0017+0055. The gray regions correspond to the wavelength range considered in each absorption feature. Shading about the model fit represents the inverse SNR for each feature ($\xi = 1/SNR$), determined from maximum likelihood estimation.
\label{fig:GI_fits}}
\end{figure*}

\begin{figure*}
\centering
\includegraphics[width=\textwidth, trim=7.0cm 0.0cm 7.0cm 0.0cm, clip]{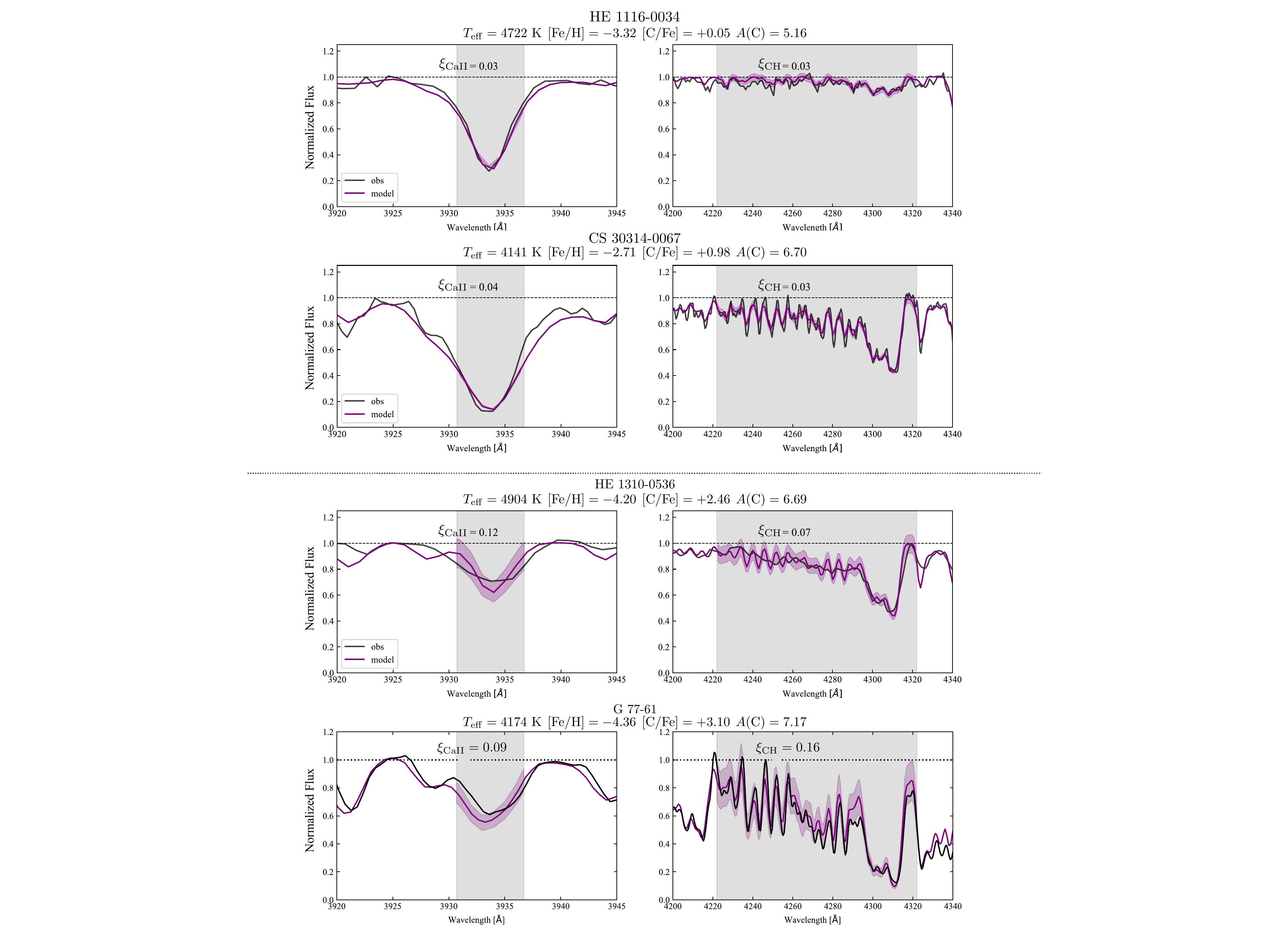}
\caption{MCMC MLE Model fits (purple) of the \ion{Ca}{2} K \textit{(left column)}, CH $G$-band \textit{(right column)}, for the Group II (HE~1116-0034, CS~30314-0067) and Group III (HE~1310-0536 and G~77-61) validation stars . The gray regions correspond to the wavelength range considered in each absorption feature. Shading about the model fit represents the the inverse SNR for each feature ($\xi = $1/SNR), determined from maximum likelihood estimation.
\label{fig:GII_GIII_fits}}
\end{figure*}

Figures~\ref{fig:classification} -- \ref{fig:GII_GIII_fits} present the fitting results for our validation stars, all of which have published high-resolution spectroscopic analyses.

\end{document}